\documentclass{elsart}
\usepackage{amsfonts}
\usepackage{amsmath}
\usepackage{amssymb}
\usepackage{graphicx}

\newcommand\openone{\leavevmode\hbox{\small1\normalsize\kern-.33em1}}
\newcommand{\captions}{\sf\caption}

\def\dmsol{\Delta m_\odot^2}
\def\dmatm{\Delta m_a^2}

\begin{document}

\begin{frontmatter}

\title{Enlarging~the~window~for~radiative~leptogenesis}

\author[Lisbon]{G. C. Branco}, \author[Lisbon]{R. Gonz\'alez Felipe}, \author[Padova]{F. R.
Joaquim}, \author[Lisbon]{B. M. Nobre}

\address[Lisbon]{Centro de F\'{\i}sica Te\'{o}rica de Part\'{\i}culas, Departamento de F\'{\i}sica,
Instituto Superior T\'{e}cnico, Av. Rovisco
Pais, 1049-001 Lisboa, Portugal}

\address[Padova]{Dipartimento di Fisica ``G. Galilei'', Universit\`a di Padova and INFN,
Sezione di Padova, Via Marzolo, 8 - I-35131 Padua - Italy}

\begin{abstract}
We investigate the scenario of resonant thermal leptogenesis, in which the
leptonic asymmetries are generated through renormalization group corrections
induced at the leptogenesis scale. In the framework of the standard model
extended by three right-handed heavy Majorana neutrinos with masses $M_1 = M_2
\ll M_3$ at some high scale, we show that the mass splitting and $CP$-violating
effects induced by renormalization group corrections can lead to values of the
$CP$ asymmetries large enough for a successful leptogenesis. In this scenario,
the low-energy neutrino oscillation data can also be easily accommodated. The
possibility of having an underlying symmetry behind the degeneracy in the
right-handed neutrino mass spectrum  is also discussed.
\end{abstract}

\end{frontmatter}

\section{Introduction}\label{sec:Introduction}

Among the viable mechanisms to explain the matter-antimatter asymmetry observed
in the universe, leptogenesis~\cite{Fukugita:1986hr} has undoubtedly become one
of the most compelling ones~\cite{Buchmuller:2005eh,Hambye:2004fn}. Indeed, its
simplicity and close connection with low-energy neutrino physics render
leptogenesis an attractive and eventually testable scenario. Unfortunately,
even in its simplest realization through the well-known seesaw
mechanism~\cite{Minkowski:1977sc}, the theory is plagued with too many
parameters. To appreciate this point, let us recall that in the framework of
the standard model (SM) extended with three heavy Majorana neutrinos $N_{i}\,
(i=1,2,3)$, the high-energy neutrino sector, characterized by the Dirac
neutrino ($m_D$) and the heavy Majorana neutrino ($M_R$) mass matrices, has
eighteen parameters. Of them, only nine combinations enter into the seesaw
effective neutrino mass matrix $m_D\,M_R^{-1}\,m_D^T\,$, thus making difficult
to establish a direct link between leptogenesis and low-energy
phenomenology~\cite{Branco:2001pq}. Furthermore, there are six $CP$-violating
phases which are physically relevant at high energies, while only three
combinations of them are potentially observable at low energies. Therefore, no
direct link between the sign of the baryon asymmetry and low-energy leptonic
$CP$ violation can be established, unless extra assumptions are introduced.

In a more economical framework~\cite{Frampton:2002qc}, where only two heavy
Majorana neutrinos are present, one is left with eleven high-energy parameters,
three of which are physical phases. But since in this case one light neutrino
is predicted to be massless, there remain only seven independent low-energy
neutrino parameters, two of which are $CP$-violating phases. Thus, additional
assumptions are usually required to completely determine the high-energy
neutrino sector from low-energy observables. Typical examples are the
introduction of texture zeros in the Yukawa matrices or the imposition of
symmetries to constrain their structure~\cite{Kaneko:2002yp}. In this respect,
the heavy Majorana neutrino masses are rather unconstrained: they can range
from the TeV region to the GUT scale, and the spectrum can be hierarchical,
quasi-degenerate or even exactly degenerate~\cite{GonzalezFelipe:2001kr}.
Despite this arbitrariness, the heavy Majorana neutrino mass scale (and,
consequently, the seesaw scale) turns out to be crucial for a successful
implementation of the leptogenesis mechanism. In particular, the standard
thermal leptogenesis scenario with hierarchical heavy Majorana neutrino masses
($M_1 \ll M_2 < M_3$) requires $M_1 \gtrsim 4 \times
10^8$~GeV~\cite{Hamaguchi:2001gw}, if $N_{1}$ is in thermal equilibrium before
it decays, or the more restrictive lower bound $M_1 \gtrsim 2 \times
10^9$~GeV~\cite{Giudice:2003jh} for a zero initial $N_{1}$ abundance. Since
this bound also determines the lowest reheating temperature allowed after
inflation, it could be problematic in supersymmetric theories due to the
overproduction of light particles like the gravitino.

Nevertheless, it is important to note that the above bounds are not model
independent in the sense that they can be avoided, if the heavy Majorana
neutrino spectrum is no longer hierarchical. Indeed, if at least two of the
$N_{i}$ are quasi-degenerate in mass, \emph{i.e.} $M_1 \simeq M_2\,$, then the
leptonic $CP$ asymmetry relevant for leptogenesis exhibits the resonant
behaviour $\varepsilon_1 \sim
M_1/(M_2-M_1)$~\cite{Pilaftsis:1997jf,Pilaftsis:2003gt}. In this case, it is
possible to show that the upper bound on the $CP$ asymmetry is independent of
the light neutrino masses and successful leptogenesis simply requires $M_{1,2}$
to be above the electroweak scale for the sphaleron interactions to be
effective. Of course, having such a degeneracy in the neutrino masses requires
a compelling justification. Not being accidental, the quasi-degeneracy may
arise, for instance, from some flavour symmetry softly broken at a high
scale~\cite{D'Ambrosio:2003wy}.

Another possibility which has been recently
explored~\cite{GonzalezFelipe:2003fi,Turzynski:2004xy} relies on the fact that
radiative effects, induced by the renormalization group (RG) running from high
to low energies, can naturally lead to a sufficiently small neutrino mass
splitting at the leptogenesis scale. In the latter case, nonvanishing and
sufficiently large $CP$ asymmetries, which are proportional to the
charged-lepton $\tau$ Yukawa coupling and weakly dependent on the heavy
Majorana neutrino mass scale, are generated. Although in the SM framework the
resulting baryon asymmetry turns out to be (by a factor of two) below the
observed value~\cite{GonzalezFelipe:2003fi}, this mechanism can be successfully
implemented in its minimal supersymmetric extension
(MSSM)~\cite{Turzynski:2004xy}. However, it is worth emphasizing that the above
results have been obtained in a minimal seesaw scenario with only two heavy
neutrinos. In such a case, low seesaw scales require extremely small Dirac
neutrino Yukawa couplings for both $N_1$ and $N_2$, in order to avoid too large
low-energy neutrino masses. One may therefore ask whether the above problems
can be overcome in a more realistic scenario where the effects of a third heavy
Majorana neutrino $N_3$ are also taken into account.

The purpose of this paper is to further investigate the scenario of radiative
leptogenesis proposed in Ref.~\cite{GonzalezFelipe:2003fi}, in which the
leptonic $CP$ asymmetries are generated through RG corrections induced at the
leptogenesis scale. In the framework of the standard model extended by the
addition of three heavy Majorana neutrinos with masses $M_1 = M_2 \ll M_3$ at
some high scale, we show that the mass splitting induced by the running of the
heavy neutrino masses can lead to values of the $CP$ asymmetries large enough
for a successful leptogenesis. In this scenario, the observed baryon asymmetry
and low-energy neutrino oscillation data can be easily reconciled. Moreover,
since the results depend very weakly on the gap between the degeneracy and
leptogenesis scales, low right-handed neutrino masses and reheating
temperatures are acceptable, thus avoiding the well-known problem of
overproduction of relic abundances in early universe. Finally, we shall also
comment on possible symmetries which could explain the degeneracy of
right-handed neutrino masses at high energies.

\section{Radiative leptogenesis}\label{sec:Radiative-effects}

In the SM extended by the addition of three right-handed neutrinos, the
relevant Yukawa and heavy Majorana neutrino mass terms in the Lagrangian are
\begin{equation}
\mathcal{L}  \propto\; \bar{\ell}_L Y_\ell\, \ell_{R}\, \phi^0 +\bar{\nu}_L
Y_\nu\, N\, \phi^0 - \frac{1}{2} N^{T} C M_R\, N + {\rm H.c.}\,, \label{Lyuk}
\end{equation}
where $\ell\,$ and $\nu$ refer to the charged-lepton and neutrino fields,
respectively (family indices are omitted); $Y_\ell$ and $Y_\nu$ are the
charged-lepton and Dirac neutrino Yukawa coupling matrices and $\phi^0$ denotes
the neutral component of the SM Higgs doublet. After integrating out the heavy
Majorana neutrinos $N$, the light neutrino mass matrix, resulting from the
seesaw mechanism, is given by
\begin{equation}\label{eq:seesaw}
\mathcal{M} = - v^2\, Y_\nu\,M_R^{-1}\,Y_\nu^T\,,\, v \equiv \langle \phi^0
\rangle\,.
\end{equation}

In the basis where $Y_\ell$ and $M_R$ are diagonal, all the parameter space can
be conveniently spanned through the parametrisation~\cite{Casas:2001sr},
\begin{equation}
    Y_\nu = \frac{1}{v}\, U\, d^{1/2}\, R\, D^{1/2}\, ,
    \label{CI_parameterization}
\end{equation}
where $d = \text{diag}\, (m_1\,,\,m_2\, e^{i \alpha},m_3\, e^{i \beta})$ and $D
= \text{diag}\, (M_1\,, M_2\,,M_3)$; $m_i$ are the light neutrino masses and
$\alpha, \beta$ are Majorana phases. The matrix $R$ is an arbitrary $3 \times
3$ complex orthogonal matrix, which can be parameterised in terms of complex
angles $\theta_i$ as
\begin{equation}
\label{R_parametrization}
R  = \left\lgroup
\begin{array}{ccc}
  c_{1}\, c_{2}\quad & s_{1}\, c_{2}\quad & s_{2} \\
  -s_{1}\, c_{3} - c_{1}\, s_{2}\, s_{3}\quad & c_{1}\, c_{3}
  - s_{1}\, s_{2}\, s_{3}\quad  & c_{2}\, s_{3} \\
  - c_{1}\, s_{2}\, c_{3} + s_{1}\, s_{3}\quad &
  - s_{1}\, s_{2}\, c_{3} - c_{1}\, s_{3}\quad &
  c_{2}\, c_{3}
\end{array}
\right\rgroup\, ,
\end{equation}
where $s_{i} \equiv \sin \theta_i$,  $c_{i} \equiv \cos \theta_i$. Finally, the
matrix $U$ is the standard Pontecorvo-Maki-Nakagawa-Sakata (PMNS) leptonic
mixing matrix, which contains the $CP$-violating Dirac phase $\delta$. It turns
out that the parametrisation (\ref{CI_parameterization}) is also particularly
convenient to disentangle the $CP$-violating phases relevant for leptogenesis
from the low-energy phases. Indeed, the combination
\begin{equation}\label{equ:Hnu}
 H \equiv Y_\nu^\dagger Y_\nu =\frac{1}{v^2}\, D^{1/2}\, R^\dagger\, |d|\, R\,
 D^{1/2}\,,
\end{equation}
which appears in physical quantities associated with leptogenesis is only
sensitive to the phases in $R$ and not to the phases $\alpha, \beta$ and
$\delta$. In terms of the matrix elements of $R$, the matrix $H$ reads as
\begin{align} \label{Hij}
  H_{ij} = \frac{\sqrt{M_i M_j}}{v^2}\, \sum_{k=1}^3\, m_k\,R^\ast_{ki}\,
  R_{kj}\,,  \quad (i,\,j=1,2,3)\,.
\end{align}

Let us now discuss how the resonant leptogenesis mechanism works in the present
framework. We assume an exact degeneracy of two heavy Majorana neutrinos, so
that $M_1=M_2\equiv M \ll M_3$ at a scale $\Lambda$, which is higher than the
decoupling scale of the heaviest neutrino $N_3$. The parameter
\begin{align}
\label{deltaN} \delta_N \equiv \frac{M_2}{M_1}-1\,,
\end{align}
quantifies the degree of degeneracy between $M_1$ and $M_2$ at lower scales.
Assuming that the interactions involving $N_{1,2}$ are in thermal equilibrium
at the time the heaviest neutrino $N_3$ decays, only the leptonic $CP$
asymmetries generated in the out-of-equilibrium decays of $N_1$ and $N_2$ will
be relevant for leptogenesis. These asymmetries are given by
\begin{equation}
    \label{CPAsymmetry}
    \varepsilon_j \simeq
    \frac{\mathrm{Im}[H^2_{21}]}{\, 16\, \pi\, H_{jj}\,\delta_N}\left(1+
    \frac{\Gamma^2_i}{4\, M^2_j\, \delta^2_N}\right)^{-1}\, ,\,
    \Gamma_i=\frac{H_{ii}\, M_i}{8\, \pi}\,, \quad i, j= 1,\,
    2\; (i \neq j)\, ,
\end{equation}
where $\Gamma_i$ are the tree-level decay widths. Notice that
Eqs.~(\ref{CPAsymmetry}) exhibit the expected resonant enhancement due to the
mixing of two nearly degenerate heavy Majorana
neutrinos~\cite{Pilaftsis:1997jf}. In the present framework, a sufficiently
small heavy neutrino mass splitting will be generated through the RG running
effects. The latter turn out to be crucial in this
case~\cite{GonzalezFelipe:2003fi,Turzynski:2004xy,Antusch:2005gp}.

The evolution of the right-handed neutrino masses and the Dirac neutrino Yukawa
matrix is given at one-loop by~\cite{Casas:1999tp}
\begin{align}
\label{RGEMRdiag}
\frac{d M_i}{dt}&=2 M_i\, H_{ii}\,,  \nonumber\\
\frac{d Y_\nu}{dt}&=\left[ T- \frac{3}{4} g_Y^2 - \frac{9}{4} g_2^2
-\frac{3}{2} \left(Y_\ell Y_\ell^\dagger-Y_\nu Y_\nu^\dagger\right) \right]
Y_\nu + Y_\nu A\,,
\end{align}
where $t \equiv \frac{1}{16\, \pi^2}\,\ln(\mu/\Lambda)$, $T = 3 \text{Tr}(Y_u
Y_u^\dagger)+3 \text{Tr}(Y_d Y_d^\dagger)+\text{Tr}(Y_\ell
Y_\ell^\dagger)+\text{Tr}(Y_\nu Y_\nu^\dagger)$; $Y_{u,d}$ are the up-quark and
down-quark Yukawa matrices and $g_{Y,2}$ are the gauge couplings. The matrix
$A$ is antihermitian with
\begin{align} \label{Rmatrix}
A_{jj}=0\,,\quad A_{jk}=\frac{M_j+M_k}{M_k-M_j}\, \text{Re}\,(H_{jk})+
i\frac{M_k-M_j}{M_j+M_k} \text{Im}\,(H_{jk}) =-\; A_{jk}^\ast\,.
\end{align}
As is clear from the above equation, to avoid the singularity of $A_{12}$ at
the degeneracy scale $\Lambda$, we must require $\text{Re}\,(H_{12})=0$. This
condition can always be guaranteed without loss of generality. Indeed, when
$M_1=M_2$ there is the freedom to rotate the right-handed neutrino fields
$N_{1,2}$ with a real orthogonal matrix that does not change $M_R\,$, but
rotates $Y_{\nu}$ to the appropriate
basis~\cite{GonzalezFelipe:2003fi,Turzynski:2004xy}. In terms of the
parametrisation (\ref{CI_parameterization}), this is equivalent to a
redefinition of the real part of the complex angle $\theta_1$. In this sense,
$\text{Re}\, \theta_1$ is no longer a free parameter, \emph{i.e.} it is
constrained by the condition $\text{Re}\,(H_{12})=0$.

At this point it is worthwhile to comment on the number of physical parameters
in the high-energy neutrino sector. From the previous analysis it is clear that
in the case of two degenerate heavy Majorana neutrinos, there remain 16
physical parameters out of 18. A similar conclusion can be easily drawn by
parameterising the Dirac neutrino Yukawa coupling matrix in the general form
$Y_\nu=V\, Y_\triangle$. Here $V$ is a unitary matrix containing 3
$CP$-violating phases and $Y_\triangle$ is a lower triangular matrix with real
diagonal entries and having in general 3 phases in the
off-diagonal~\cite{Branco:2001pq}. It is then straightforward to show that the
requirement $\text{Re}\,(H_{12})=0$ leads to a constraint on one of the
physical phases of $Y_\triangle$. This should not come as a surprise. Indeed,
the correct counting of independent $CP$-violating phases always requires that
one chooses an appropriate basis.

We now proceed with the estimate of the radiatively induced $CP$ asymmetries at
the leptogenesis scale $\mu \approx M$~\cite{GonzalezFelipe:2003fi}. At a given
scale $\mu$, the degeneracy parameter $\delta_N$ is approximately given by
\begin{align}
  \delta_N(t) \simeq 2 (H_{22}-H_{11})\, t\, .
  \label{deltaNinduced}
\end{align}
From this simple expression we see that the lifting of the $N_1-N_2$ degeneracy
requires $H_{22} \neq H_{11}$. Moreover, even if at the degeneracy scale
$\Lambda$ one has $\text{Re}\,(H_{12})=0$, a nonvanishing real part will be
generated by quantum corrections. From Eqs.~(\ref{RGEMRdiag}) and
(\ref{Rmatrix}) we find
\begin{align}
  \label{reH12}
  \text{Re} [H_{21}(t)] &\simeq -\frac{3}{2}\, y^2_\tau\, \text{Re}
  \left[(Y_\nu)^\ast_{31}\, (Y_\nu)_{32}\right]\, t\, .
\end{align}
Thus, neglecting the RG running of $\text{Im} (H_{21})$, one has for the
$CP$-violating part appearing in the leptonic asymmetries,
\begin{align}
\label{imH12s}
  \text{Im} [H^2_{21}(t)] &\simeq -3\, y^2_\tau\, \text{Im} [H_{21}(0)]\,
  \text{Re} \left[(Y_\nu)^\ast_{31}\, (Y_\nu)_{32}\right]\, t\, ,
\end{align}
where $y_\tau$ is the $\tau$ Yukawa coupling. In terms of the elements of $R$,
the quantity $\text{Re} \left[(Y_\nu)^\ast_{31}\, (Y_\nu)_{32}\right]$ reads as
\begin{align}
  \text{Re} \left[(Y_\nu)^\ast_{31}\, (Y_\nu)_{32}\right]  = \frac{M}{v^2}\,
   \sum_{i,j=1}^3\,\sqrt{m_i\, m_j}\, \text{Re}
  \left[ R^\ast_{i1}\, R_{j2}\, U^\ast_{3i}\, U_{3j} \right]\, .
\end{align}
Here, the dependence of the radiatively induced $\text{Re}\,(H_{12})$ on the
low-energy parameters is evident through the presence of the light neutrino
masses $m_i$ and the elements of the third row of the mixing matrix $U$. We
also notice that only a small dependence on the mixing parameter $U_{e3}$ is
expected.

Substituting Eqs.~(\ref{deltaNinduced}) and (\ref{imH12s}) into
Eq.~(\ref{CPAsymmetry}) we obtain the following expressions for the leptonic
$CP$ asymmetries,
\begin{equation}
\label{epsigen}%
    \varepsilon_{1,2} \simeq \varepsilon^0_{1,2}\, (1 + D_{2,1})^{-1}\,,
\end{equation}
where $ \varepsilon^0_{1,2}$ are the uncorrected $CP$ asymmetries and $D_{1,2}$
are correction factors which include the effects of the heavy Majorana decay
widths,
\begin{align}
    \label{CPAsymmetry_uncorrected}
    \varepsilon^0_j &\simeq \frac{3y^2_\tau}{32\, \pi}
    \frac{\text{Im} (H_{21})\,
    \text{Re} \left[(Y_\nu)^\ast_{31}\, (Y_\nu)_{32}\right]}
     {H_{jj}(H_{22}-H_{11})}\, , \\
    D_j &\simeq \frac{1}{(32\, \pi)^2}\frac{H^2_{jj}}{(H_{22} - H_{11})^2\, t^2} \, .
    \label{Dj}
\end{align}

We note that the leptonic $CP$ asymmetries do not depend explicitly on the
heaviest mass $M_3\,$. Moreover, if the corrections due to the inclusion of the
decay widths in the propagators are negligible, \emph{i.e.} $D_j \ll 1$, the
asymmetries are also independent of the mass $M$ of the two lightest
right-handed Majorana neutrinos.

The expressions for the leptonic $CP$ asymmetries in terms of the parameters
$\theta_i$ are quite long. Therefore, we will consider some interesting
limiting cases for which simple analytical expressions can be obtained and the
viability of the present mechanism is readily demonstrated. The simplest cases
are clearly those with a single non-vanishing parameter $\theta_i$. Since
$\theta_1 = \theta_2 = 0$ and $\theta_1 = \theta_3 = 0$ imply $H_{12} = 0$,
these would lead to vanishing asymmetries. Thus, we are left with the case
$\theta_2 = \theta_3 = 0$, which we consider next.

\subsection{The case $\theta_2=\theta_3=0$}

Since in this case the condition $\text{Re}\,(H_{12})=0$ implies that the
complex angle $\theta_1$ is purely imaginary, \emph{i.e.} $\theta_1= i
\omega_1$, the elements of the matrix $H$ relevant for leptogenesis are simply
given by
\begin{align} \label{Hcase1}
  H_{11} & = \frac{M}{v^2}\, \left(m_1\, \cosh^2 \omega_1 + m_2\, \sinh^2
  \omega_1 \right)\, , \nonumber\\
  H_{22} & = \frac{M}{v^2}\, \left(m_1\, \sinh^2 \omega_1 + m_2\, \cosh^2
  \omega_1 \right)\, ,\\
  H_{12} &= i \frac{M}{2\, v^2}\, \sinh(2\, \omega_1)\,
  (m_1 + m_2)\,,\nonumber
\end{align}
so that
\begin{equation}
  H_{11}-H_{22} = \frac{M}{v^2}\, (m_1 - m_2)\, .
\end{equation}
It is interesting to note that there is a direct connection between the induced
heavy Majorana mass splitting parameter $\delta_N$ (cf.
Eq.~(\ref{deltaNinduced})) and the low-energy neutrino mass spectrum.

Below the degeneracy scale $\Lambda$, $H_{12}$ develops a real part
proportional to
\begin{equation}
  \text{Re}[(Y_\mu)^\ast_{31}\, (Y_\nu)_{32}] \simeq \frac{M}{v^2}\,
  \sqrt{m_1 \, m_2}\, \text{Re}\, (U^\ast_{32}\, U_{31})\, .
\end{equation}
According to Eq.~(\ref{CPAsymmetry_uncorrected}), the uncorrected leptonic
$CP$-asymmetries $\varepsilon^0_1$ and $\varepsilon^0_2$ are then given by
\begin{align}
\label{eps0}%
  \varepsilon^0_1 &\simeq \frac{3\, y^2_\tau}{64\, \pi}\,
  \frac{(m_1 + m_2)\, \sqrt{m_1\, m_2}\, \sinh(2\, \omega_1)\,
  \text{Re}\, (U^\ast_{32}\, U_{31})}{(m_1 - m_2)(m_1\, \cosh^2 \omega_1 +
  m_2\, \sinh^2 \omega_1)}\, , \nonumber\\ \\
\varepsilon^0_2  &\simeq \frac{3\, y^2_\tau}{64\, \pi}\, \frac{(m_1 + m_2)\,
\sqrt{m_1\, m_2}\, \sinh(2\, \omega_1)\,
  \text{Re}\, (U^\ast_{32}\, U_{31})}{(m_1 - m_2)(m_1\, \sinh^2 \omega_1 +
  m_2\, \cosh^2 \omega_1)}\, .\nonumber
\end{align}
The corrections due to the inclusion of the heavy Majorana neutrino decay
widths are obtained from Eq.~(\ref{Dj}),
\begin{align}\label{Dcoeff}
  D_1 & \simeq \left[\frac{\pi}{2}\, \frac{m_1\, \cosh^2 \omega_1 + m_2\, \sinh^2 \omega_1}
  {(m_2 - m_1)\, \ln \left(\Lambda/M\right)}
  \right]^2\,  , \nonumber\\ \\
  D_2 & \simeq \left[\frac{\pi}{2}\, \frac{m_1\, \sinh^2 \omega_1 + m_2\, \cosh^2 \omega_1}
  {(m_2 - m_1)\, \ln \left(\Lambda/M\right)}
  \right]^2 \,. \nonumber
\end{align}
It is interesting to analyse how the $CP$-asymmetries $\varepsilon_{1,2}$
behave in some limiting cases. First, we notice that for $m_1=0$, the
quantities $\varepsilon_{1,2}^0$ vanish and no lepton asymmetry is generated.
Consequently, in this case a lower bound on $m_1$ is expected in order to
reproduce the observed baryon asymmetry.\footnote{This is no longer true in the
most general case $\theta_{2,3} \neq 0$, as confirmed by the results of the
next section.} On the other hand, being $\varepsilon_{1,2}^0$ proportional to
$(m_2-m_1)^{-1}$, one could expect an enhancement in the limit $m_1 \simeq
m_2$. However, in this case the corrections due to the heavy Majorana decay
widths become important, since $D_{1,2} \propto (m_2-m_1)^{-2}$. From
Eqs.~(\ref{epsigen}), (\ref{eps0}) and (\ref{Dcoeff}) one gets that
$\varepsilon_{1,2} \propto m_2-m_1$ which explains the suppression of the
baryon asymmetry for $m_1 \simeq m_2$. This is the case of quasi-degenerate or
inverted-hierarchical light neutrinos. In conclusion, when
$\theta_2=\theta_3=0$ and the light neutrinos are hierarchical, one can expect
an interval of intermediate values of the lightest mass $m_1$ for which the
radiative leptogenesis mechanism could lead to a sufficient baryon asymmetry.

In contrast to what happens in the standard thermal leptogenesis scenario, the
$CP$-asymmetries in the present framework depend explicitly on the PMNS mixing
matrix $U$. In the simple case under analysis, this dependence appears through
the combination $\text{Re}\, (U^\ast_{32}\, U_{31})$ as shown in
Eqs.~(\ref{eps0}). Thus, the final value of the baryon asymmetry will depend on
the particular values of $U_{e3}$ and the low-energy $CP$-violating phases
$\alpha, \beta$ and $\delta$ (as well as on the neutrino mass-squared
differences and the solar and atmospheric mixing angles, which however we
assume already fixed by the data).

\subsection{The case $m_1=0$}

In spite of all the major experimental advances in the measurement of the
neutrino mixing parameters, no information about leptonic $CP$ violation is
available yet. While the Dirac phase $\delta$ can be potentially measured in
future neutrino oscillation experiments, the only hope for probing the Majorana
phases $\alpha$ and $\beta$ seems to reside in neutrinoless double $\beta$
decay processes, which if observed, could provide only a single constraint on
these phases~\cite{Pascoli:2005zb}. In practical terms, this means that one
cannot perform a perfect experiment to completely determine the effective
neutrino mass matrix $\mathcal{M}$ from input data. Nevertheless, if this
matrix appears to be constrained so that the number of independent parameters
is reduced, then it is reasonable to require this constraint to be weak-basis
independent. One example of such a constraint is the condition $\det
\mathcal{M} = 0$~\cite{Branco:2002ie}. In this case, a massless neutrino is
predicted and the spectrum is fully hierarchical. As already mentioned, a
similar situation is verified in a minimal seesaw framework with only two-right
handed heavy Majorana neutrinos, which automatically leads to $m_1=0$. It is
therefore of interest to investigate whether the present mechanism is
compatible with the above light neutrino mass spectrum.

First we notice that in the case that $m_1=0$, the so-called minimal seesaw
scenario, which corresponds to the two heavy Majorana neutrino limit, can be
obtained by setting $\theta_1=i \omega_1,\,\theta_2 = \pi/2$ and $\theta_3=0$.
Therefore, to present our analytical results we consider the simplest
generalisation of the latter case by letting $\theta_2 \equiv \omega_2$ to be
an arbitrary real parameter. We then find
\begin{align} \label{Hcase2}
  H_{11} & = \frac{M}{v^2}\, \left(m_2\, \sinh^2 \omega_1 +
  m_3\, \cosh^2 \omega_1\, \sin^2 \omega_2 \right)\,, \nonumber\\
  H_{22} & = \frac{M}{v^2}\, \left(m_2\, \cosh^2 \omega_1 +
  m_3\, \sinh^2 \omega_1\, \sin^2 \omega_2 \right)\,,\\
  H_{12} &= i\,\frac{M}{2 v^2} \sinh(2\, \omega_1) \left(m_2 +
  m_3\, \sin^2 \omega_2 \right)\,. \nonumber
\end{align}
Moreover,
\begin{equation}
  H_{11} - H_{22} = \frac{M}{v^2}\, (-m_2 + m_3\, \sin^2 \omega_2)
\end{equation}
and
\begin{equation}
  \text{Re}[(Y_\mu)^\ast_{31}\, (Y_\nu)_{32}] \simeq - \frac{M}{v^2}\,
  \sqrt{m_2\, m_3}\,\sin\, \omega_2 \, \text{Re}(U^\ast_{32}\, U_{33})\,.
\end{equation}
From the above equations it is clear that, contrarily to what happened in the
previous case where the radiatively generated $\delta_N$ and $\text{Re}
(H_{12})$ depended exclusively on low-energy parameters, these two quantities
depend now also on the structure of the orthogonal matrix $R$ through the
parameter $\omega_2$.

The leptonic $CP$-asymmetries $\varepsilon^0_i$ are given in this case by
\begin{align} \label{eps0m10}
  \varepsilon^0_1 &\simeq \frac{3\, y^2_\tau}{64\, \pi}\,
  \frac{\sqrt{m_2\, m_3}\, \sin \omega_2\, \sinh (2\, \omega_1)
  (m_2 + m_3\, \sin^2 \omega_2)\, \text{Re}(U^\ast_{32}\, U_{33})}{(-m_2 +
    m_3\, \sin^2 \omega_2)
  (m_2\, \sinh^2 \omega_1 + m_3\, \cosh^2 \omega_1\, \sin^2\,
  \omega_2)}\,,\nonumber\\ \\
  \varepsilon^0_2 &\simeq \frac{3\, y^2_\tau}{64\, \pi}\,
  \frac{\sqrt{m_2\, m_3}\, \sin \omega_2\, \sinh (2\, \omega_1)
  (m_2 + m_3\, \sin^2 \omega_2)\, \text{Re}(U^\ast_{32}\, U_{33}) }{(-m_2
  + m_3\, \sin^2 \omega_2)
  (m_2\, \cosh^2 \omega_1 + m_3\, \sinh^2 \omega_1\, \sin^2\,
  \omega_2)}\,,\nonumber
\end{align}
and the factors $D_i$ read
\begin{align}\label{Dcoeffm10}
  D_1 & \simeq \left[\frac{\pi}{2}\,\frac{m_2\, \sinh^2 \omega_1 + m_3\, \cosh^2 \omega_1\,
  \sin^2 \omega_2}{(-m_2 + m_3\, \sin^2 \omega_2)\,
  \ln\left(\Lambda/M\right)}\right]^2\,, \nonumber\\ \\
  D_2 & \simeq \left[\frac{\pi}{2}\,\frac{m_2\, \cosh^2 \omega_1 + m_3\, \sinh^2 \omega_1\,
  \sin^2 \omega_2}{(-m_2 + m_3\, \sin^2 \omega_2)\,
  \ln \left(\Lambda/M\right)}\right]^2\,.\nonumber
\end{align}
As expected, when $\omega_2 = \pi/2$ the results of the minimal seesaw scenario
considered in \cite{GonzalezFelipe:2003fi,Turzynski:2004xy} are recovered.
Thus, the new contributions coming from the mixing with the heaviest Majorana
neutrino $N_3$ turn out to be crucial in this case for the mechanism to be
viable.

From Eqs.~(\ref{eps0m10}) and (\ref{Dcoeffm10}) it is clear that the leptonic
$CP$-asymmetries $\varepsilon_i$ vanish when
\begin{align}\label{critw2}
    \sin^2 \omega_2 =\frac{m_2}{m_3} =
    \left(\frac{\dmsol}{\dmatm}\right)^{1/2}\,,
\end{align}
where $\dmsol$ and $\dmatm$ are the mass-squared differences measured in solar
and atmospheric neutrino oscillation experiments, respectively. For values of
$\omega_2$ close to the above value, the contribution of the coefficients $D_i$
becomes relevant. We also note that for small values of $\omega_2$ one has
$\varepsilon_1/\varepsilon_2 \simeq \coth^2 \omega_1 >1$.

\section{Numerical results and discussion}

The most recent WMAP results and BBN analysis of the primordial deuterium
abundance imply~\cite{Spergel:2003cb}
\begin{equation}
\eta_{B}=\frac{n_B}{n_\gamma}=(6.1\pm0.3)\times10^{-10}\,,\label{BAU}
\end{equation}
for the baryon-to-photon ratio of number densities. In the leptogenesis
framework, once a lepton asymmetry has been generated by the out-of-equilibrium
decays of the heavy Majorana neutrinos, it will be converted into a baryon
asymmetry by non-perturbative sphaleron interactions. The efficiency in
producing the asymmetry is controlled by the parameters
\begin{equation}
K_i=\frac{\tilde{m_i}}{m_\ast}\, , \quad \tilde{m}_i =
    \frac{v^2\, H_{ii}}{M_i}\,,
    \label{decay_parameters_i}
\end{equation}
where $m_\ast \simeq 10^{-3}\, \text{eV}$ is the so-called equilibrium neutrino
mass. The resulting baryon asymmetry can be estimated as
\begin{equation}
  \eta_B \simeq -10^{-2}\, (\kappa_1\, \varepsilon_1 + \kappa_2\,
  \varepsilon_2)\,,
\end{equation}
where $\kappa_i < 1$ are the efficiency factors, which account for the washout
effects. An accurate computation of these factors requires the solution of the
relevant Boltzmann equations. In our numerical calculations we make use of the
Boltzmann equations derived in Ref.~\cite{Pilaftsis:2003gt}, which are
appropriate for resonant leptogenesis and, therefore, suitable to the cases
considered here. We also remark that leptogenesis in the present framework
always occurs in a strong washout regime. Indeed, from Eqs.~(\ref{Hcase1}),
(\ref{Hcase2}) and (\ref{decay_parameters_i}) it follows that
\begin{align}
K_1+K_2 > K_\odot \equiv \frac{(\dmsol)^{1/2}}{m_\ast} \simeq 9\,.
\end{align}
In this situation, the simple decay-plus-inverse-decay picture is applicable
and the final baryon asymmetry is essentially independent of the initial
conditions~\cite{diBari}.

Our numerical computations proceed as follows. We start at $\mu=M_Z$ with the
best-fit values for the solar and atmospheric neutrino oscillation
paramaters~\cite{Strumia:2005tc}:
\begin{align}\label{data}
    \tan^2\theta_{12}&=0.45\,,\, \dmsol=8.0\times 10^{-5}\,\text{eV}^2\,,
    \nonumber\\
    \tan^2\theta_{23}&=1.0\,,\,\,\, \dmatm=2.5\times 10^{-3}\,\text{eV}^2\,.
\end{align}
For a given set of $U_{e3}$, $m_1$ and $CP$-violating phases, the low-energy
effective neutrino mass matrix
$\mathcal{M}=U^\dag\,\text{diag}(m_1,m_2,m_3)\,U^\ast$ is constructed. For the
hierarchical (inverted-hierarchical) neutrino mass spectrum the lightest
neutrino mass $m_1$ ($m_3$) is an input parameter. The two remaining masses are
$m_2^2=m_1^2+\dmsol\,,\,m_3^2=m_1^2+\dmsol+\dmatm$ for a hierarchical spectrum
and $m_1^2=m_3^2+\dmatm-\dmsol\,,\,m_2^2=m_3^2+\dmatm$ for an inverted
hierarchy. For a particular choice of the input parameters, the RG equations
for the neutrino masses and mixing angles are solved up to the degeneracy
scale, which we consider to be $\Lambda \simeq 10^{16}$~GeV. At this stage, we
do not consider the running effects due to the Dirac neutrino Yukawa couplings
above the mass of the lightest heavy Majorana neutrino. We then define $Y_\nu$
at the scale $\Lambda$ as in Eq.~(\ref{CI_parameterization}), using a specific
pattern for $R$ and fixing the values of $M_1=M_2=M$. The value of $M_3>M$ is
fixed by requiring the largest Dirac neutrino Yukawa coupling to be equal to
the top-quark Yukawa coupling $y_t$. For the simplest viable scenario where $R$
is parameterised by $\theta_1=i\,\omega_1$ and $\theta_2=\theta_3=0$, this is
equivalent to $M_3=y_t^2\,v^2/m_3$.

All the couplings and masses are subsequently evolved down to the scale
$\mu=M$, considering also the decoupling of $N_3$. At this scale the baryon
asymmetry is computed as described at the beginning of this section. Obviously,
the two heavy Majorana neutrinos $N_1$ and $N_2$ are no longer degenerate at
$\mu=M$ due to radiative effects. Moreover, a non-trivial $CP$-violating part
is generated due to the running of $Y_\nu$. We also evolve the effective
neutrino mass operator from $M$ down to $M_Z$ in order to check whether the
inclusion of $Y_\nu$ and threshold corrections in the top-down running affects
the values of the neutrino parameters initially considered. If so, the
parameters at the degeneracy scale $\Lambda$ are accordingly changed to achieve
convergence.

\begin{figure}
$$\begin{array}{cc}
  \!\!\!\!\includegraphics[width=7cm]{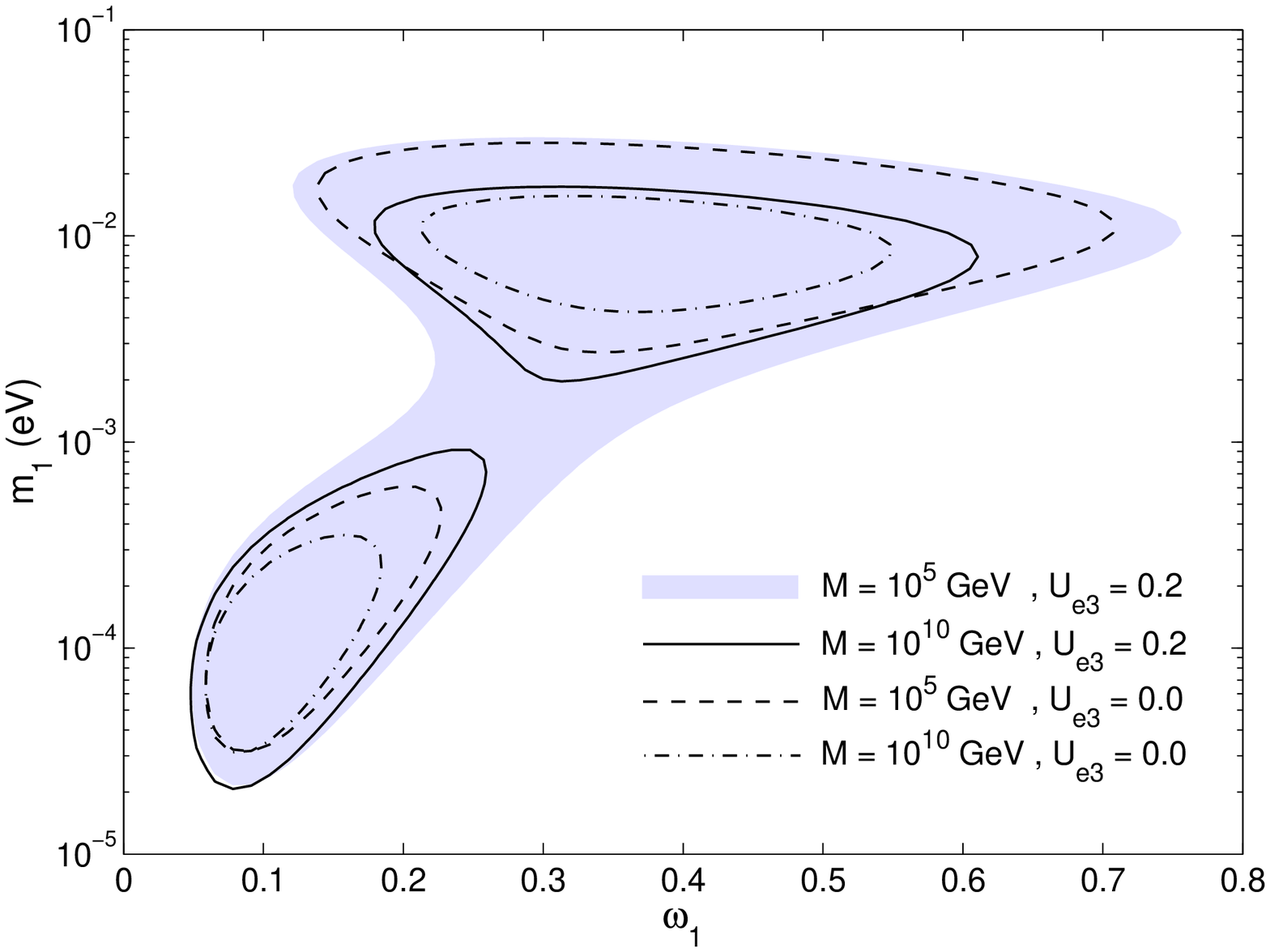} &
  \includegraphics[width=7cm]{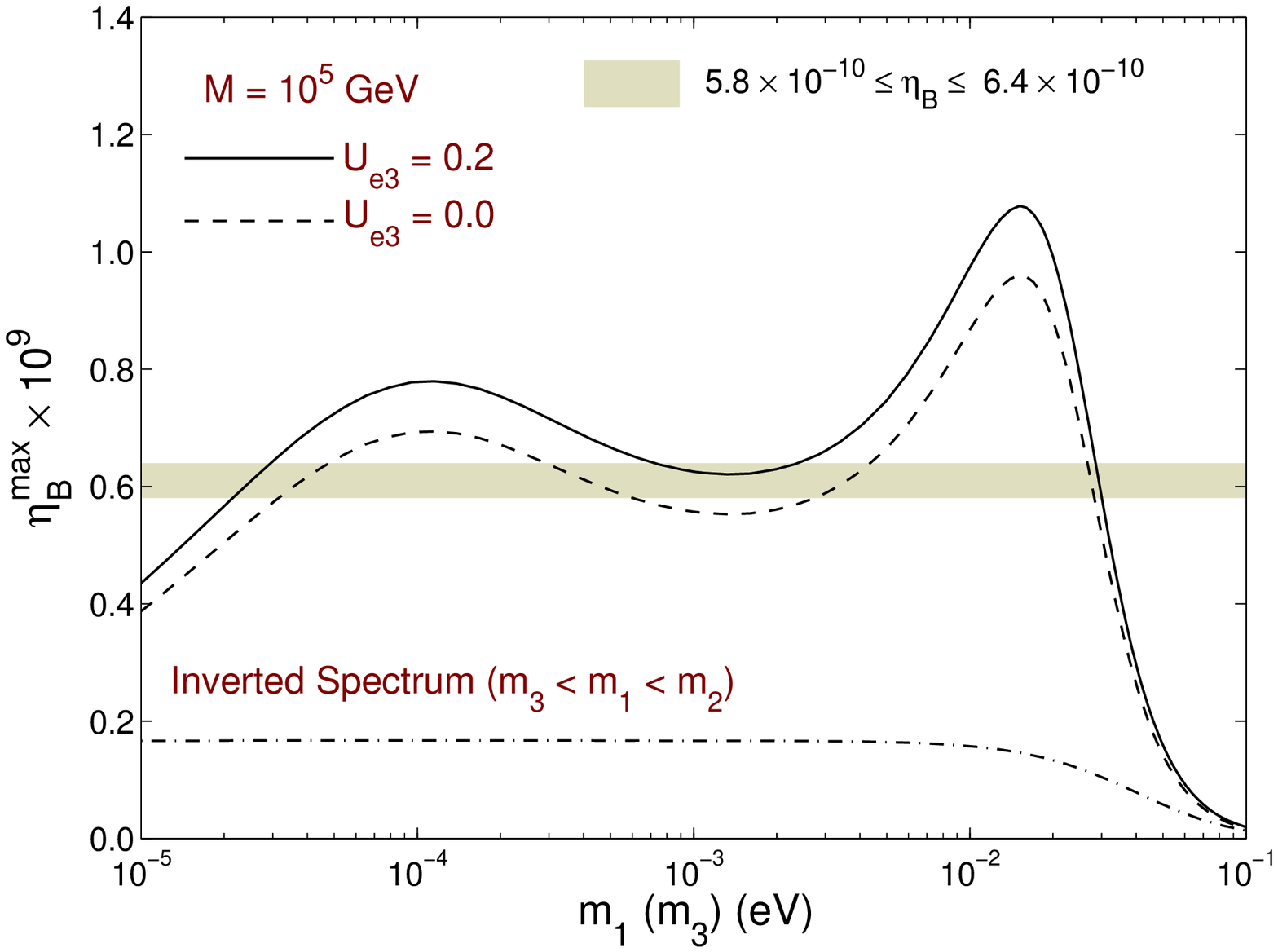}
\end{array}
$$
\captions{The baryon asymmetry as a function of $\omega_1$ in the case
$\theta_2=\theta_3=0$ for $U_{e3}=0, 0.2$. On the left, the regions in the
$(\omega_1,m_1)$-plane where the value of $\eta_B \geq 5.8 \times 10^{-10}$.
The results are presented for $M=10^5,\,10^{10}$~GeV. The maximal value of
$\eta_B$ as a function of the lightest neutrino mass $m_1$ is shown in the
right plot for the same values of $U_{e3}\,$ and $M=10^5$~GeV. The dash-dotted
line corresponds to $\eta_B^{\text{max}}$ as a function of $m_3$ for an
inverted neutrino mass spectrum $m_3<m_1<m_2$.}\label{fig1}
\end{figure}

In Fig.~\ref{fig1} we present the results of our numerical analysis for the
case $\theta_2=\theta_3=0$. The plot on the left shows the allowed region in
the $(\omega_1,m_1)$-plane where $\eta_B$ can be larger than the lower bound
given in Eq.~(\ref{BAU}), \emph{i.e.} $\eta_B \geq 5.8 \times 10^{-10}$. The
contours are given for $U_{e3}= 0$ and $U_{e3}= 0.2$. The filled region was
obtained following the full numerical procedure and considering $M=10^5$~GeV.
Changing the scale $M$ to $10^{10}$~GeV does not alter the results
significantly, as can be seen from the figure. This interesting feature of our
scenario can be understood by noting that the uncorrected $CP$-asymmetries
given in Eqs.~(\ref{eps0}) are independent of $M$ and $\Lambda$. From the
analysis of the plot we conclude that in this simple scenario radiative
leptogenesis is compatible with the observed baryon asymmetry, provided that
the lightest neutrino mass is in the range $2\times 10^{-5}~\text{eV}\lesssim
m_1 \lesssim 3\times 10^{-2}~\text{eV}$.

The maximal value of $\eta_B$ as a function of $m_1$ is shown in the right plot
of Fig.~\ref{fig1} for the same values of $U_{e3}\,$ and $M=10^5$~GeV. It is
interesting to note that for $m_1 \lesssim 10^{-3}$~eV, the contributions
coming from the decay widths are negligible, $D_{1,2} \ll 1$. This explains the
small dependence of the results on the mass scale $M$. Moreover, in this region
$\varepsilon_1 \gg \varepsilon_2$ and $K_2 \sim K_\odot \gg K_1$, so that the
washout is dominated by the inverse decays of $N_2$. On the other hand, for
$m_1 \simeq 10^{-2}$~eV one has $D_1 \simeq D_2 \approx 1$ and the decay width
corrections start to become relevant.

Finally, when $m_1 \simeq m_2\,$, which corresponds to quasi-degenerate or
inverted-hierarchical light neutrinos, it follows directly from
Eqs.~(\ref{eps0}) and (\ref{Dcoeff}) that $\varepsilon_1 \simeq \varepsilon_2$
and $D_1 \simeq D_2 \gg 1$. As a consequence, the generated leptonic $CP$
asymmetries are suppressed by $\dmsol$ and grow with $\ln^2 (\Lambda/M)$,
\begin{equation}
  \varepsilon_1 \simeq \varepsilon_2 \simeq \frac{3 y^2_\tau}{16 \pi^3}\,
  \frac{\dmsol\, \mathrm{sinh}(2 \omega_1)\,
  \mathrm{Re}\,(U^\ast_{32}\, U_{31})}{m_1^2\, (2\, \mathrm{cosh}^2
  \omega_1-1)^3}\, \ln^2 \left(\frac{\Lambda}{M}\right)\, .
\end{equation}

In this situation, the expression for the baryon asymmetry is approximately
given by
\begin{equation}
  \eta_B \simeq 4 \times 10^{-12} \left(\frac{\dmsol}{8 \times 10^{-5}\,\mathrm{eV}^2}\right)\,
  \left(\frac{0.05\,\mathrm{eV}}{m_1}\right)^3\,
  \frac{\, \mathrm{sinh}(2 \omega_1)\,
  \mathrm{Re}\,(U^\ast_{32}\, U_{31})}{(2\, \mathrm{cosh}^2
  \omega_1-1)^4}\, \ln^2 \left(\frac{\Lambda}{M}\right)\, .
\end{equation}
In particular, for an inverted hierarchy with $m_1 \approx \sqrt{\dmatm} \simeq
0.05\,\mathrm{eV}\,$, one can show that $\eta_B$ is maximal when $m_3 = 0$,
$\omega_1 \simeq 0.3$ and $\mathrm{Re}\, (U^\ast_{32}\, U_{31}) \simeq 1/4$.
Thus, $\eta_B$ is bounded by
\begin{equation}
   \eta_B \lesssim 3 \times 10^{-13} \ln^2 \left(\frac{\Lambda}{M}\right)\, .
\end{equation}
When $M=10^5$~GeV and $\Lambda=10^{16}$~GeV, this upper bound corresponds to
the plateau shown in the right plot of Fig.~\ref{fig1}, obtained numerically
for an inverted neutrino mass spectrum.

\begin{figure}
$$\begin{array}{cc}
  \!\!\!\!\!\includegraphics[width=7cm]{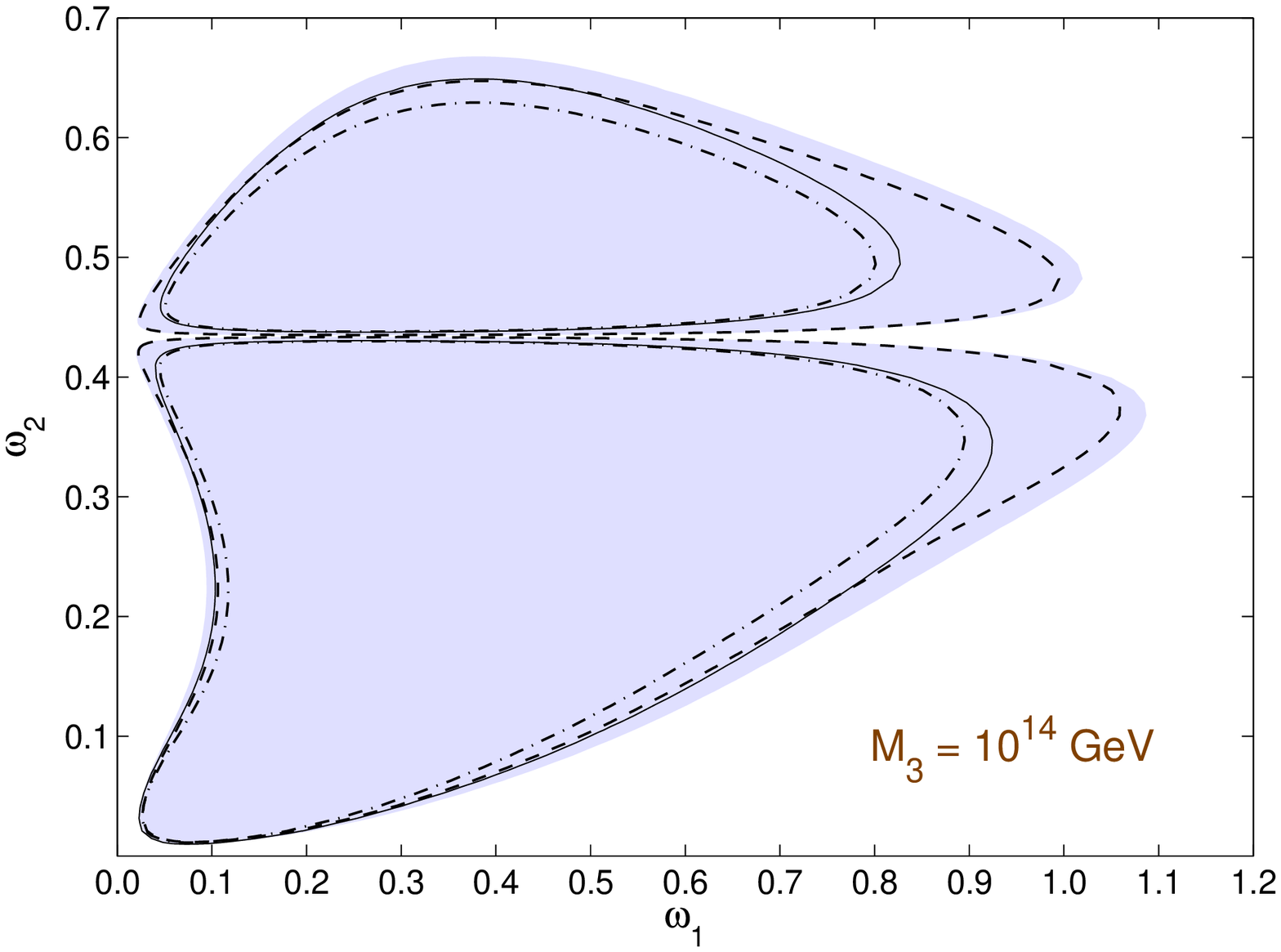} &
  \includegraphics[width=7cm]{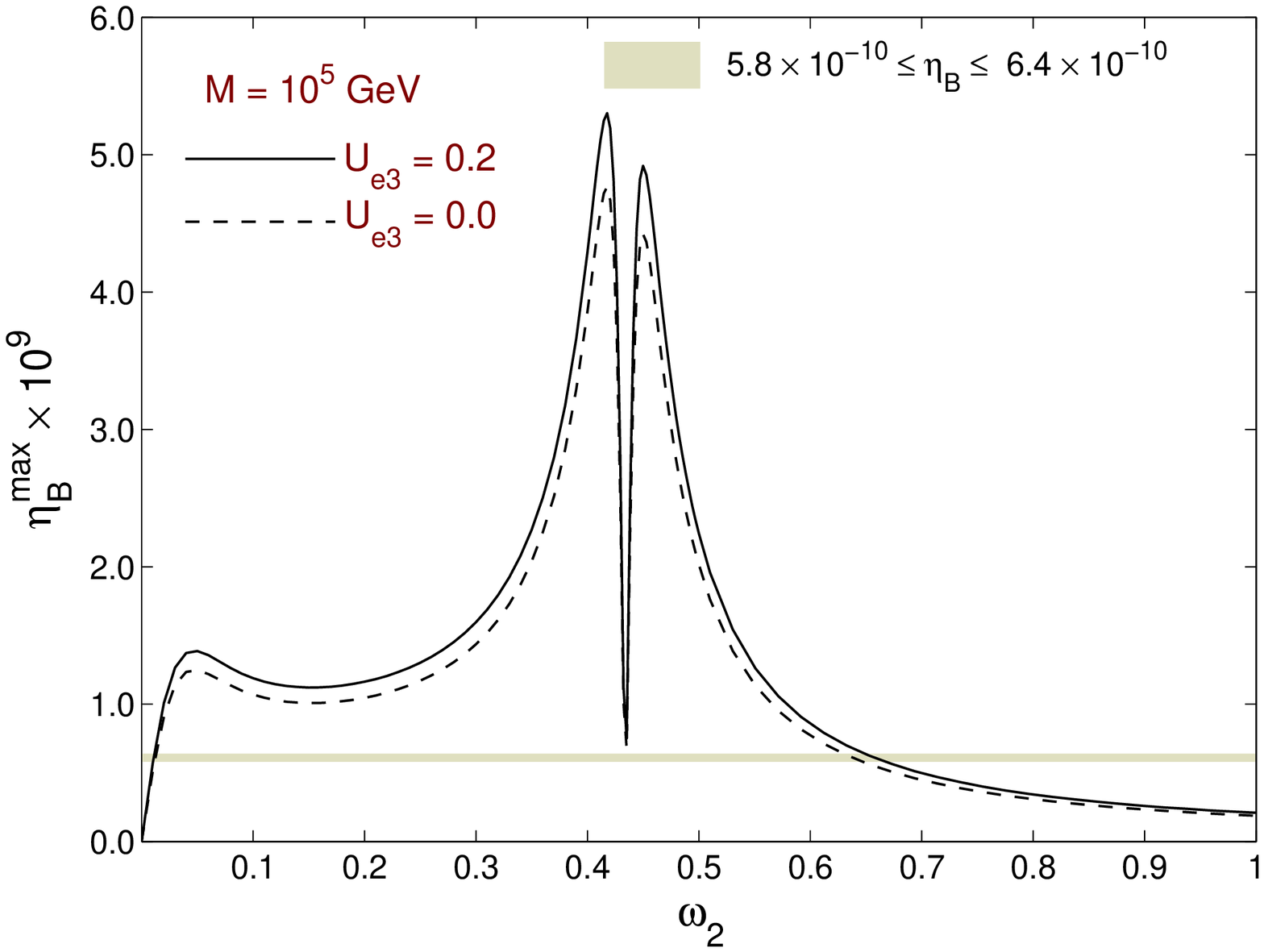}
\end{array}
$$
\captions{The baryon asymmetry as a function of $\omega_1$ and $\omega_2$ in
the case $m_1=0\,,\,\theta_3=0$ for $U_{e3}=0, 0.2$; $M=10^5, 10^{10}$~GeV and
$M_3=10^{14}~$GeV. The line and colour schemes are the same as used in
Fig.~\ref{fig1}. On the left plot, the region in the
$(\omega_1,\omega_2)$-plane where the value of $\eta_B \geq 5.8 \times
10^{-10}$ is shown. The maximal value of $\eta_B$ as a function of $\omega_2$
is shown in the right plot for the same values of $U_{e3}\,$and
$M=10^5$~GeV.}\label{fig2}
\end{figure}

Similar plots are shown in Fig.~\ref{fig2} for the case $m_1=0$, $\theta_1=i
\omega_1$, $\theta_2=w_2$ and $\theta_3=0$. On the left, we present the allowed
regions in the $(\omega_1,\omega_2)$-plane where the lower bound $\eta_B = 5.8
\times 10^{-10}$ can be attained. As in Fig.~\ref{fig1}, we take $U_{e3}=0,\,
0.2$, $M=10^5, 10^{10}$~GeV and $M_3=10^{14}~$GeV. As expected, there are two
distinct allowed regions separated by the line corresponding to the value of
$\omega_2$ given in Eq.~(\ref{critw2}), $\omega_2 \simeq 0.44$, where the
leptonic asymmetries vanish. As can be seen from the figure, for values of
$\omega_2$ close to the above value there is a clear dependence on the mass
parameter $M$. This has to do with the fact that in that region the corrections
due to the $N_{1,2}$ decay widths are significant, $D_{1,2} \gtrsim 1$. The
maximal value of $\eta_B$ as a function of $\omega_2$ is shown in the right
plot for the same values of $U_{e3}$ and $M=10^5$~GeV.

From the previous analysis we conclude that radiative leptogenesis in the
framework of the SM extended with 3 heavy Majorana neutrinos is compatible with
a hierarchical light neutrino mass spectrum. As anticipated by the analytical
calculations, the small sensitivity of $\eta_B$ to the value of $U_{e3}\,$ is
confirmed by our numerical results. This can be readily seen by comparing the
curves for $U_{e3}=0$ and $0.2$ in both Figs.~\ref{fig1} and \ref{fig2}.

\section{On a symmetry behind the heavy Majorana neutrino degeneracy}

An essential aspect of the resonant leptogenesis framework is the requirement
of having two quasi-degenerate heavy neutrino masses. In the radiative
leptogenesis scenario, this quasi-degeneracy is generated by RG corrections,
starting from a situation where there is an exact degeneracy between the heavy
Majorana neutrino masses. This is, in our opinion, one of the most natural
frameworks for the implementation of resonant leptogenesis\footnote{See
Refs.~\cite{Pilaftsis:1997jf,West:2004me} for some works where the question of
heavy Majorana neutrino degeneracy has also been addressed.}. It is therefore
important to investigate whether this can be achieved by an underlying symmetry
principle.

First, we should remark that, in order for our mechanism to work, any symmetry
leading to exact heavy Majorana neutrino degeneracy between $N_1$ and $N_2$ has
to be such that $H_{11}-H_{22}\neq 0$. This is crucial in order to radiatively
generate the mass splitting $\delta_N$, as can be seen from
Eq.~(\ref{deltaNinduced}). Moreover, Eqs.~(\ref{reH12}) and (\ref{imH12s})
require $\text{Re} \left[(Y_\nu)^\ast_{31}\, (Y_\nu)_{32}\right]\neq 0$ and
$\text{Im} [H_{12}]\neq 0$, otherwise the $CP$-violating effects needed for
leptogenesis will be highly suppressed.

We now address the question of whether the degeneracy in the heavy Majorana
neutrino mass spectrum could reflect the presence of an underlying symmetry  at
a high-energy scale. We shall briefly comment on two possible scenarios, based
on simple discrete or Abelian symmetries. The more ambitious program of
extending these symmetries to the quark and lepton sectors of the full theory
is beyond the scope of this paper and will be presented elsewhere.

Let us assume that there is a $Z_3\times S_3$ symmetry, under which the
right-handed neutrino fields $N_i$ transform in the following way:
\begin{align}
Z_3 : \; N_i \rightarrow P_{ij} N_j\,, \quad S_3 : \; N_1 \rightarrow N_2\,,\,
N_2 \rightarrow N_3\,,\, N_3 \rightarrow N_1\,,
\end{align}
where the matrix $P$, which defines the transformation properties of the $N_i$
under the $Z_3$ symmetry, is given by\footnote{It is interesting to note that
this $Z_3$ symmetry is the minimal discrete symmetry which leads to a scenario
with extended flavour democracy~\cite{Branco:2001hn,Akhmedov:2000yt}.}
\begin{align}
 P=i\, \omega^* W\,, \quad
 W=\frac{1}{3} \left(\begin{array}{ccc}
                      \omega & 1 & 1 \\
                      1 &\; \omega \;& 1 \\
                      1 & 1 & \omega \\
                      \end{array}
                \right)\,,\quad \omega=e^{i 2\pi/3}\,.
\end{align}
It can be readily verified that the most general Majorana mass term which is
consistent with the above symmetry is $M_R = M_0\, \Delta$, where $M_0$ is a
high mass scale and $\Delta$ is the $3\times 3$ democratic matrix,
$\Delta_{ij}=1$. We now assume that the $Z_3\times S_3$ symmetry is softly
broken into $S_3$, so that the right-handed Majorana mass matrix has the form
\begin{align}
M_R = M_0\,(\Delta + \epsilon\, \openone)\,,
\end{align}
where $ |\epsilon| \ll 1$. One can easily verify that the eigenvalues of this
matrix lead to required spectrum: $M_1=M_2=|\epsilon| M_0$ and $M_3 \simeq 3
M_0$. The parameter $\epsilon$ reflects here the hierarchy between the scales
$M_3$ and $M_{1,2}$. Note that assuming $ |\epsilon| \ll 1$ is natural in the
't Hooft sense~\cite{tHooft}, since in the limit $\epsilon \rightarrow 0$ the
matrix $M_R$ acquires a larger symmetry, namely, $Z_3\times S_3$.

Next we consider another possible explanation for the degeneracy in $M_i\,$,
based on Abelian symmetries. In fact, one of the most popular schemes
considered to explain the fermion mass and mixing patterns is the
Froggatt-Nielsen mechanism~\cite{Froggatt:1978nt} with spontaneously broken
Abelian flavour symmetries~\cite{Altarelli:2004za}. Such flavour symmetries are
assumed to be broken by $\langle X\rangle/M_\ast=\epsilon\ll 1$, where $X$ is a
scalar field and $M_\ast$ is the fundamental mass scale of the theory. In order
to try to explain the required heavy Majorana neutrino mass spectrum, we
consider in the context of a supersymmetric theory a model with two Abelian
flavour symmetries $U(1)_X\times U(1)_{X^\prime}$. The two scalar fields $X$
and $X^\prime$ are assumed to have charges $Q(X)=(-1,-1)$ and
$Q(X^\prime)=(0,1)$ under $U(1)_X\times U(1)_{X^\prime}$. In this case, the
effective superpotential contains the following nonrenormalizable terms for the
heavy Majorana neutrino masses
\begin{align}
\label{WN}%
W_N=c_{ij}\,M_{B-L}\left(\frac{X}{M_\ast}\right)^{x_{ij}}\left(\frac{X^\prime}
{M_\ast}\right)^{x_{ij}-x_{ij}^\prime}\!\!N_iN_j\;,\;
x_{ij}^{(\prime)}=n_i^{(\prime)}+n_j^{(\prime)}\,,
\end{align}
where $n_i^{(\prime)}$ is the charge of $N_i$ under the $U(1)_{X(X^\prime)}$
symmetry. The $c_{ij}$ are order one coefficients not determined by the flavour
symmetry and $M_{B-L}$ is the typical $B-L$ breaking scale. After spontaneous
breaking of the $U(1)$, the heavy Majorana neutrino mass matrix is given by
\begin{align}
\label{MRU1}%
(M_R)_{ij}=c_{ij}\,M_{B-L}\,\epsilon_1^{x_{ij}}\epsilon_2^{x_{ij}-x_{ij}^\prime}\;,
\;\epsilon_1=\frac{\langle X \rangle}{M_\ast}\,,\,\epsilon_2=\frac{\langle
X^\prime \rangle}{M_\ast}\,,
\end{align}
with $\epsilon_{1,2}\ll 1$. Since the appearance of nonrenormalizable terms
with negative powers of the superfields $X$ and $X^\prime$ is forbidden by the
holomorphycity of the superpotential, the $U(1)$ charges $n_i^{(\prime)}$ have
to be such that $x_{ij}\geq 0$ and $x_{ij}-x_{ij}^\prime\geq0$. Otherwise,
$c_{ij}$ must be set to zero (holomorphic zeros)~\cite{Leurer:1992wg}. This
property can be used to justify a heavy Majorana mass spectrum of the type:
$M_1=M_2 \ll M_3 \simeq M_{B-L}$. Indeed, it is easy to see that imposing the
conditions
\begin{align}
\label{conditions}%
&x_{ij}\geq 0\,\wedge\,x_{ij}-x_{ij}^\prime\geq 0\;,\;(i,j)=(1,2),(3,3)\,,\nonumber\\
&x_{ij}<0\,\vee\,x_{ij}-x_{ij}^\prime<0\;,\;(i,j)\neq (1,2),(3,3)\,,
\end{align}
one obtains the following structure
\begin{align}
\label{MRU1s} &(M_R)_{12}=c_{12}\,\epsilon_1^{x_{12}^{}}
\epsilon_2^{x_{12}-x_{12}^\prime}M_{B-L}\;,\;
(M_R)_{33}=c_{33}\,\epsilon_1^{x_{33}^{}}
\epsilon_2^{x_{33}-x_{33}^\prime}M_{B-L} \,,\nonumber\\
&(M_R)_{ij}=0\,,\,(i,j)\neq(1,2),(3,3)\,.
\end{align}
which leads to  $M_1=M_2=|(M_R)_{12}|$ and $M_3=|(M_R)_{33}|$. There is however
a caveat on this approach. It is well known that in these schemes, besides the
usual canonical terms $N_i^\dag N_i\,$, the K\"{a}hler potential receives
nonrenormalizable contributions involving powers of $X/M_\ast$ and
$X^\prime/M_\ast\,$. As emphasized in Refs.~\cite{Dudas:1995yu}, these extra
terms may fill the supersymmetric zeros corresponding to negative powers of the
scalar fields in the superpotential. This is a consequence of the superfield
redefinitions which bring back the K\"{a}hler potential to the canonical form. As a
result, the superpotential couplings get modified~\cite{Espinosa:2004ya}. In
the present case, one can show that, after the $U(1)$ spontaneous symmetry
breaking, the K\"{a}hler potential reads
\begin{align}
\label{Kcorre}%
K=N_i^\dag\,\mathcal{C}_{ij}\,N_j\;,\;
\mathcal{C}_{ij}=\left(\delta_{ij}+k_{ij}\,\epsilon_1^{|a_{ij}^{}|}
\epsilon_2^{|a^\prime_{ij}|}\right)\,,
\end{align}
where $k_{ij}$ are coefficients, $a_{ij}=n_j-n_i$ and
$a^\prime_{ij}=n_j-n_i+n_i^\prime-n^\prime_j\,$.

Obviously, the transformation which redefines the superfields $N_i$ to the
canonical basis will depend on the choice of the charges $n_{i}$ and
$n_{i}^\prime$. For illustration, let us take the following set of $n_i$ and
$n_i^\prime$ charges
\begin{align}
\label{charges}%
n_i=(2,-1,0)\;,\;n_i^\prime=(3,-4,0)\,,
\end{align}
which obey the conditions given in Eq.~(\ref{conditions}), and let us assume
$\epsilon_1 \simeq \epsilon_2 \equiv \epsilon$. From the above charge
configuration it follows that the uncorrected $M_R$ leads to $M_1=M_2\simeq
\epsilon^3\,M_{B-L}$ and $M_3 \simeq M_{B-L}$. One can show that the
redefinition of the heavy neutrino superfields performed to recover the
canonical form of the K\"{a}hler potential lifts the degeneracy between $N_1$ and
$N_2$ with a corresponding $\delta_N =M_2/M_1-1 \sim \epsilon^3$. Therefore, in
this case the radiative leptogenesis framework would make sense only if the RG
corrections to $\delta_N$ (cf. Eq.~(\ref{deltaNinduced})) are larger than
$\epsilon^3$. Clearly, this will depend on the size of the Dirac neutrino
Yukawa couplings. To conclude, it is worth emphasizing that if one invokes this
kind of $U(1)$ flavour symmetries to explain degenerate or quasi-degenerate
spectra, as it is in the case of resonant leptogenesis, these effects should be
properly taken into account.

\section{Conclusions}

We have presented an appealing and economical scenario of resonant
leptogenesis, based on the radiative generation of the leptonic $CP$
asymmetries. In particular, we have studied the mechanism of radiative
leptogenesis~\cite{GonzalezFelipe:2003fi} in the more general $3\times 3$ SM
seesaw framework with a heavy Majorana neutrino mass spectrum $M_1 \simeq M_2
\ll M_3$. We have shown that even for simple flavour structures of the Dirac
neutrino Yukawa coupling matrix, one can successfully generate the cosmological
baryon asymmetry and, simultaneously, accommodate the low-energy neutrino data.
The key ingredients for the viability of the mechanism are the heavy Majorana
mass splitting and the $CP$-violating effects induced at the leptogenesis scale
by renormalization group corrections. As far as leptogenesis is concerned, our
conclusions are quite independent of the specific values of the heavy Majorana
mass scales $M$ and $M_3$, as well as of the degeneracy scale $\Lambda$. We
have also seen that the mechanism works in a wide region of the low-energy
neutrino parameter space. In contrast with the minimal seesaw scenario with
only two heavy Majorana neutrinos, we have concluded that the present framework
is compatible with a fully hierarchical light neutrino mass spectrum.
Furthermore, from the simple limiting cases considered, an upper bound on the
lightest neutrino mass $m_1 \lesssim 0.03$~eV was obtained. Obviously, this
bound is expected to get modified if one considers non-minimal structures for
the neutrino Yukawa coupling matrix.

We have also presented a brief discussion on possible symmetries which could
lead to an exact mass degeneracy between $N_1$ and $N_2$ at a high-energy
scale. For instance, the soft breaking of a specific $Z_3\times S_3$ symmetry
to $S_3$ by a small parameter $\epsilon$ naturally leads to a heavy Majorana
mass spectrum of the type $M_{1,2}=\epsilon M_3$. Alternatively, flavour
structures based on $U(1)$ Abelian symmetries can also explain such a
degeneracy. However, as it was stressed, the application of such symmetries to
explain exact or quasi-degenerate mass spectra should be done with care.
Indeed, one should properly take into account the corrections which appear when
the K\"ahler potential is brought to its canonical form by a redefinition of
the heavy Majorana neutrino fields.

\section*{Acknowledgements}
We are grateful to Pasquale Di Bari for useful comments and enlightening
discussions. F.R.J. thanks Andrea Brignole and Ferruccio Feruglio for useful
discussions and CFTP (IST, Lisbon) for its kind hospitality during the final
stage of this work.

This work was supported by {\em Funda\c{c}\~{a}o para a Ci\^{e}ncia e a Tecnologia} (FCT,
Portugal) through the projects POCTI / FNU/ 44409/ 2002 and PDCT/
FP/FNU/50250/2003. R.G.F. and F.R.J. have been supported by FCT under the
grants SFRH/BPD/1549/2000 and SFRH/BPD/14473/2003, respectively. B. M. N. is
supported by a fellowship from CFTP and the FCT grant SFRH/BD/995/2000.

\end{document}